\documentclass[iop]{emulateapj}

\usepackage{natbib}
\usepackage{graphicx}
\usepackage{subfigure}

\usepackage{latexsym,amssymb}
\usepackage{amsmath}
\usepackage{ucs}
\usepackage[utf8x]{inputenc}
\usepackage{color}

\newcommand{\kms}{km\,s$^{-1}$}
\newcommand{\Msun}{M$_\odot$}

\slugcomment{Accepted for publication in ApJ Letters}
\shortauthors{Tsatsi et al.}
\shorttitle{A new channel for KDC formation}

\begin{document}

\title{A new channel for the Formation of Kinematically Decoupled Cores\\
 in Early-type galaxies}

\author{Athanasia Tsatsi\altaffilmark{1,*}, Andrea V. Macci\`o\altaffilmark{1}, Glenn van de Ven\altaffilmark{1}, and Benjamin P. Moster\altaffilmark{2}}

\affil{\altaffilmark{1} Max-Planck-Institut f{\"u}r Astronomie, K\"onigstuhl 17, 69117 Heidelberg, Germany}
\affil{\altaffilmark{2} Kavli Institute for Cosmology, University of Cambridge, Madingley Road, Cambridge CB3 0HA, UK}

\email{*tsatsi@mpia.de}

\begin{abstract}
  We present the formation of a Kinematically Decoupled Core (KDC) in an elliptical galaxy, resulting from a major merger simulation of two disk galaxies. We show that although the two progenitor galaxies are initially following a prograde orbit, strong reactive forces during the merger can cause a short-lived change of their orbital spin; the two progenitors follow a retrograde orbit right before their final coalescence. This results in a central kinematic decoupling and the formation of a large-scale ($\sim$2 kpc radius) counter-rotating core (CRC) at the center of the final elliptical-like merger remnant (\ensuremath{M_*=1.3 \times10^{ 11 } $\Msun$ }), while its outer parts keep the rotation direction of the initial orbital spin. The stellar velocity dispersion distribution of the merger remnant galaxy exhibits two symmetrical off-centered peaks, comparable to the observed ``2-$\sigma$ galaxies''. The KDC/CRC consists mainly of old, pre-merger population stars (older than 5 Gyr), remaining prominent in the center of the galaxy for more than 2 Gyr after the coalescence of its progenitors. Its properties are consistent with KDCs observed in massive elliptical galaxies. This new channel for the formation of KDCs from prograde mergers is in addition to previously known formation scenarios from retrograde mergers and can help towards explaining the substantial fraction of KDCs observed in early-type galaxies.
 
\end{abstract}


\section{Introduction}

Early-type galaxies (ETGs) are the end-products of complex assembly and evolutionary processes that determine their shape and dynamical structure. Signatures of such past processes in present-day ETGs are likely to be in the form of peculiar kinematic subsystems that reside in their central regions. Such subsystems are called Kinematically Decoupled Cores (KDCs) and they are defined as central stellar components with distinct kinematic properties from those of the main body of the galaxy \citep[e.g.][]{McDermid_2006, Krajnovic_2011, Toloba_2014}.

KDCs were first discovered using one-dimensional long-slit spectroscopic observations of the stellar kinematics of ETGs \citep{Efstathiou_1982, Bender_1988, Franx&Illingworth_1988}. More recently, integral-field unit spectroscopic surveys such as SAURON \citep{Bacon_2001}, ATLAS\textsuperscript{3D} \citep{Cappellari_2011}, or CALIFA \citep{Sanchez_2012}, being able to provide full two-dimensional observations of the stellar kinematics, have favored the detection of KDCs and revealed that a substantial fraction of ETGs in the nearby universe show kinematic decoupling in their central regions. This fraction ranges in different surveys, depending mainly on technical and sample-selection biases.

  Notably, the fraction of ETGs that host KDCs in the SAURON sample of 48 E+S0 galaxies \citep{deZeeuw_2002} is substantially high, especially in the centers of slow-rotating ETGs: 8 out of the 12  slow rotators ($\sim67$\%) from the main survey host a KDC \citep{Emsellem_2007}. In the ATLAS\textsuperscript{3D} volume-limited sample of 260 ETGs, this fraction is 47\% \citep{Krajnovic_2011}. The KDCs found  in slow rotators are typically ``old and large", with stellar populations that show little or no age differences with their host galaxy (older than 8 Gyr) and sizes larger than 1 kpc \citep{McDermid_2006, Kuntschner_2010}.

KDCs are also detected in fast-rotating ETGs: 25\% of fast rotators from the main SAURON survey host KDCs. This type of KDCs are typically ``young and compact", with stellar populations younger than 5 Gyr and sizes less than a few hundred parsecs \citep{McDermid_2006}.

 We note that these fractions establish a lower limit to the true fraction of ETGs with kinematically decoupled regions, considering projection effects, the fact that young and compact KDCs are subject to technical or observational biases \citep[e.g.][]{McDermid_2006}, while many ETGs with resolved KDCs in their centers are subject to different classifications throughout the literature \citep[e.g. $2\sigma$-galaxies, see][]{Krajnovic_2011}.
 
While a consensus is reached about the prominent existence of KDCs in luminous ETGs, the physical processes and the rate at which they are formed are still poorly understood. Young and compact KDCs in fast rotators might have formed via star-formation in situ. According to this scenario, the stellar component of the KDC is formed in initially kinematically misaligned gaseous regions, probably originating from externally accreted gas or unequal mass merging \citep{Hernquist&Barnes1991}, where the orientation of the merging orbit defines the orientation of rotation of the resulting KDC. Following this line of thought, \cite{Balcells&Quinn1990} suggested that counter-rotating cores can only result from retrograde mergers.

        However, this scenario could not hold for the large and old KDCs found in slow rotators, whose stellar population was probably formed at the same epoch as the main body of the galaxy. In this case, processes such as gas accretion or accretion of low-mass stellar systems are more likely to affect the outer parts of the galaxy and can not be consistent with observations that show no color gradients between the KDC and the surrounding galaxy \citep{Carollo_1997}.

The most plausible formation scenario that could explain the similarity of the stellar content of the KDC and the main body of the galaxy is major merging. This scenario has been confirmed in simulations \citep[e.g.][]{Bois_2010, Bois_2011}, resulting in elliptical-like and slow-rotating merger remnants hosting KDCs only when the two progenitor galaxies were initially following retrograde merger orbits. 
However, observations indicate a lower limit to the true rate of occurrence of KDCs in ETGs which can not be explained only by retrograde mergers, pointing to the need of additional KDC formation scenarios.

Here we show that, a KDC might as well result from an initially prograde major merger. The kinematic decoupling in the center of the final elliptical-like merger remnant can result from a short-lived change of the orbital spin of the two progenitor galaxies right after their second encounter. This new channel for the formation of KDCs might serve as an additional mechanism that can help towards explaining their observed rate of occurrence in ETGs.

\section{Simulation parameters}

The simulation we use is described in \cite{Moster_2011}. It was performed using the TreeSPH-code GADGET-2 \citep{Springel_2005}, including star formation and supernova feedback. The two progenitor disk galaxies are identical and they are composed of a cold gaseous disk, a stellar disk and a stellar bulge, which are embedded in a dark-matter and hot-gas halo.

The gaseous and the stellar disk of each progenitor galaxy have exponential surface brightness profiles and they are rotationally supported, while the spherical stellar bulge follows a \cite{Hernquist_1990} profile, and is initially non-rotating\footnote{We note that these initial structural properties of the progenitors are influenced by their close interaction, i.e. they develop bulge rotation, bars and spiral arms in the first few hundred Myr of the simulation.}. The dark matter halo has a \citeauthor{Hernquist_1990} profile and a spin parameter consistent with cosmological simulations \citep{Maccio_2008}. The hot gaseous halo follows the $\beta$ -profile \citep{Cavaliere_FuscoFemiano1976} and is rotating around the spin axis of the disk \citep[see][for a more detailed description of the galaxy model]{Moster_2011}.

The stellar mass of each progenitor is \ensuremath{M_*=5 \times
 10^{ 10 } $\Msun$ } and the bulge-to-disk ratio was chosen to be \ensuremath{B/D=0.22}. The mass of the cold gaseous disk is \ensuremath{M_{g,cold}=1.2 \times 10^{ 10 } $\Msun$ }, such that the gas fraction in the disk is 23\%. The virial mass of the dark matter halo is \ensuremath{M_{dm}=1.1 \times 10^{ 12 } $\Msun$ }, while the mass of the hot gaseous halo is \ensuremath{M_{g,hot}=1.1 \times 10^{ 11 } $\Msun$ }. The softening length is 100 pc for stellar, 400 pc for dark matter and 140 pc for gas particles.

The two progenitors are initially employed in a nearly unbound prograde parabolic orbit, with an eccentricity of $e=0.95$ and a pericentric distance of $\ensuremath{r_{p_1}}=13.6$ kpc. Such an orbit is representative for the most common major mergers in $\Lambda$CDM cosmology \citep{Khochfar_Burkert2006}.
The two galaxies have an initial separation of \ensuremath{d_{start}} = 250 kpc. The orbital and the rotation spin of the first galaxy are aligned, while the spin axis of the second galaxy is inclined by $\theta$=$30\,^{\circ}$ with respect to the orbital plane. The simulation lasts for 5 Gyr, such that the remnant elliptical galaxy is fully relaxed.

\section{Merger Remnant}

\subsection{Structure of the merger remnant}
 In order to connect the orbital and mass distribution of our simulated galaxy with observable properties, we create two-dimensional mock stellar mass maps as follows. Stellar particles are projected such that the galaxy is seen edge-on with respect to the initial orbital plane of the merger. Particles are then binned on a regular grid centered on the baryonic center of mass of the galaxy. We adopt a distance of 20 Mpc, so that 1 arcsec corresponds roughly to 0.1 kpc. Our grid has a size of 20x20 kpc, covering approximately twice the half-mass radius (\ensuremath{r_h}) of our galaxy, and a pixel size of 0.075 kpc, so that it corresponds to the spatial resolution of the IRAC camera of the Spitzer Space Telescope \citep{Fazio_2004}. 
 
We parametrize the galaxy's projected stellar mass distribution using the Multi-Gaussian Expansion (MGE) model \citep{Monnet_1992, Emsellem_1994}, as implemented by \cite{Cappellari_2002}.
The intrinsic shape of the remnant's stellar particle distribution is parametrized using an iterative method to obtain the best fitting ellipsoid to the distribution and to extract the eigenvalues of the mass tensor inside this ellipsoid \citep{Maccio_2008}. The intermediate-to-long and short-to-long axes ratios that we retrieve are p=0.88 and q=0.54, respectively,  and the average projected ellipticity of the remnant is $\epsilon=0.49$, estimated within 2 $\ensuremath{r_h}$.

  \subsection{Kinematics of the merger remnant}
  Stellar particles are projected along the chosen viewing angle and binned on a regular 20x20 kpc grid centered on the baryonic center of mass of the galaxy. In order to mimic real integral-field spectroscopic data, the pixel size of 0.1 kpc corresponds, at the adopted distance of 20 Mpc, approximately to the spatial resolution of the SAURON spectrograph \citep{Bacon_2001}. The bulk velocity of the galaxy is estimated within a sphere of 50 kpc around the center and subtracted from all particle velocities. Then we extract the mass-weighted stellar line-of-sight mean velocity and velocity dispersion for every pixel. The extracted kinematic maps are spatially binned using the 2D Voronoi binning method \citep{Cappellari_2003}, based on a minimum number of particles per pixel in the map. Signal corresponds to the number of particles per pixel and we adopt Poisson noise, such that our signal-to-noise ratio per bin (\ensuremath{SN_{bin}}) should correspond approximately to a target value $\ensuremath{SN_{T}\sim30}$.

 
  \begin{figure*}
 \centering
   \includegraphics[trim = 0cm 0cm 0cm 0cm, clip, width=18cm]{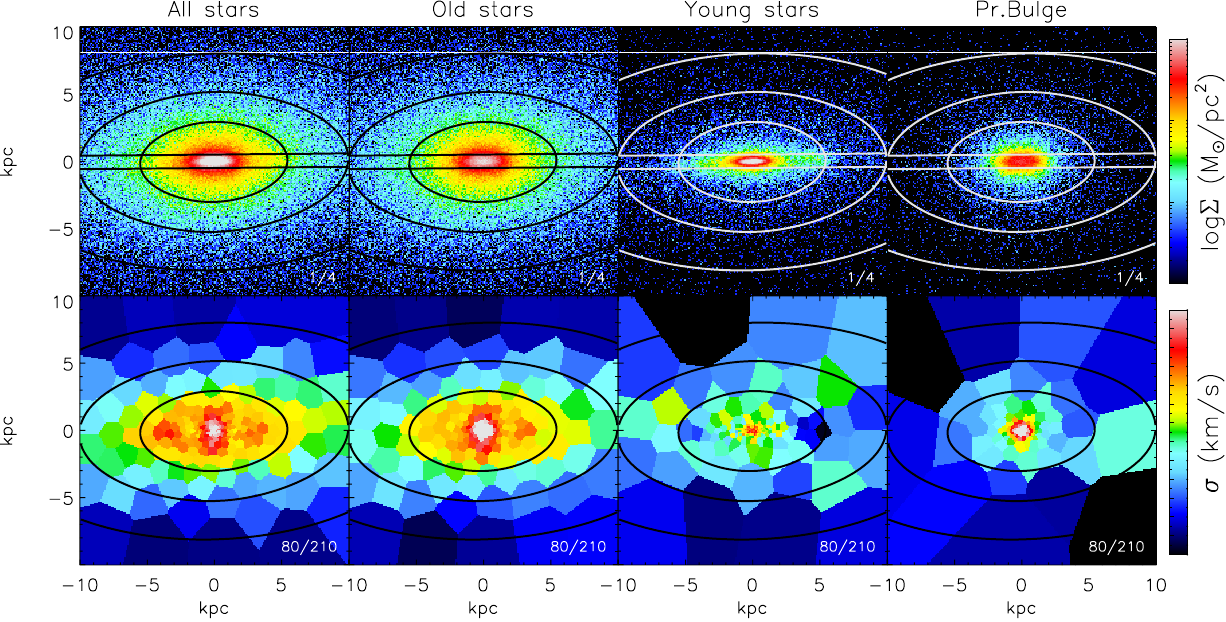}
\caption{From left to right: Component analysis of the merger remnant's stellar population: all stars (old+young), old stars (\textgreater 5 Gyr), young stars (\textless 5 Gyr), and ``Pr.Bulge'' stellar population. The latter are particles that initially formed the bulge of the progenitor galaxies. Top row shows the 2D maps of the line-of-sight projected stellar mass surface density ($\Sigma$) and bottom row shows velocity dispersion ($\sigma$) for every component. Overplotted are the MGE best-fitting contours of the total projected stellar mass, as a reference. The remnant is seen edge-on with respect to the orbital plane of the two progenitors.}
   \label{fig:figure1}
 \end{figure*}
 
 
  \begin{figure*}
 \centering
   \includegraphics[trim = 0cm 0cm 0cm 0cm, clip, width=18cm]{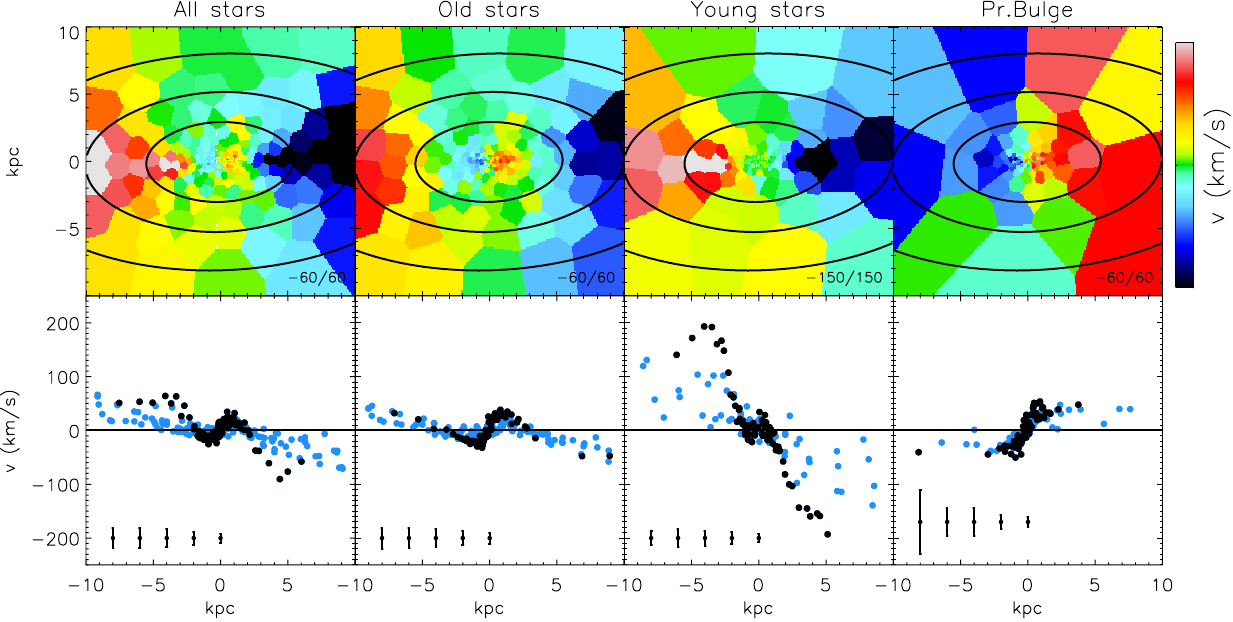}
\caption{Line-of-sight projected stellar rotation for every component of Figure~\ref{fig:figure1}. The top row shows the 2D maps of the line-of-sight projected stellar mean velocity (v) for every component. The bottom row shows the stellar rotation curve extracted from the 2D velocity maps: black points correspond to the mean velocity per bin inside a slit of 1 kpc width along the apparent major photometric axis (the position of the slit is shown at the top row figures of Figure~\ref{fig:figure1}). Blue points correspond to bins outside the slit. Typical average errors are shown at the bottom. }
   \label{fig:figure2}
 \end{figure*}
 

We also use a simple logarithmic function inferred from CALIFA data \citep{Husemann_2013} to construct mock velocity errors of our binned kinematic data:
\begin{equation}
\delta\upsilon=\ensuremath{5\,SN_{T}(1+1.4\log{N_{pix}})/SN_{bin}} \mbox{,     \kms}
\end{equation}
where \ensuremath{N_{pix}} is the number of pixels per bin.

 For the purpose of this work, we divide the stellar particles of the remnant into 4 different components: ``old stars", which are stars that initially were part of the progenitors' stellar material, (ages $\textgreater$ 5 Gyr), ``young stars", which were formed during the merger (ages $\textless$ 5 Gyr) and ``all stars", which is the total stellar content of the merger remnant. We also track the stellar particles in the remnant that initially formed the bulges of the two progenitor galaxies. These particles form the ``Pr.Bulge" (Progenitor bulge) stars. The projected two-dimensional stellar mass and the stellar kinematics for every stellar component are shown in Figures~\ref{fig:figure1} and \ref{fig:figure2}.
 
\begin{figure}
\center
\includegraphics[trim = 0cm 0cm 0cm 0cm, clip, width=8.5cm]{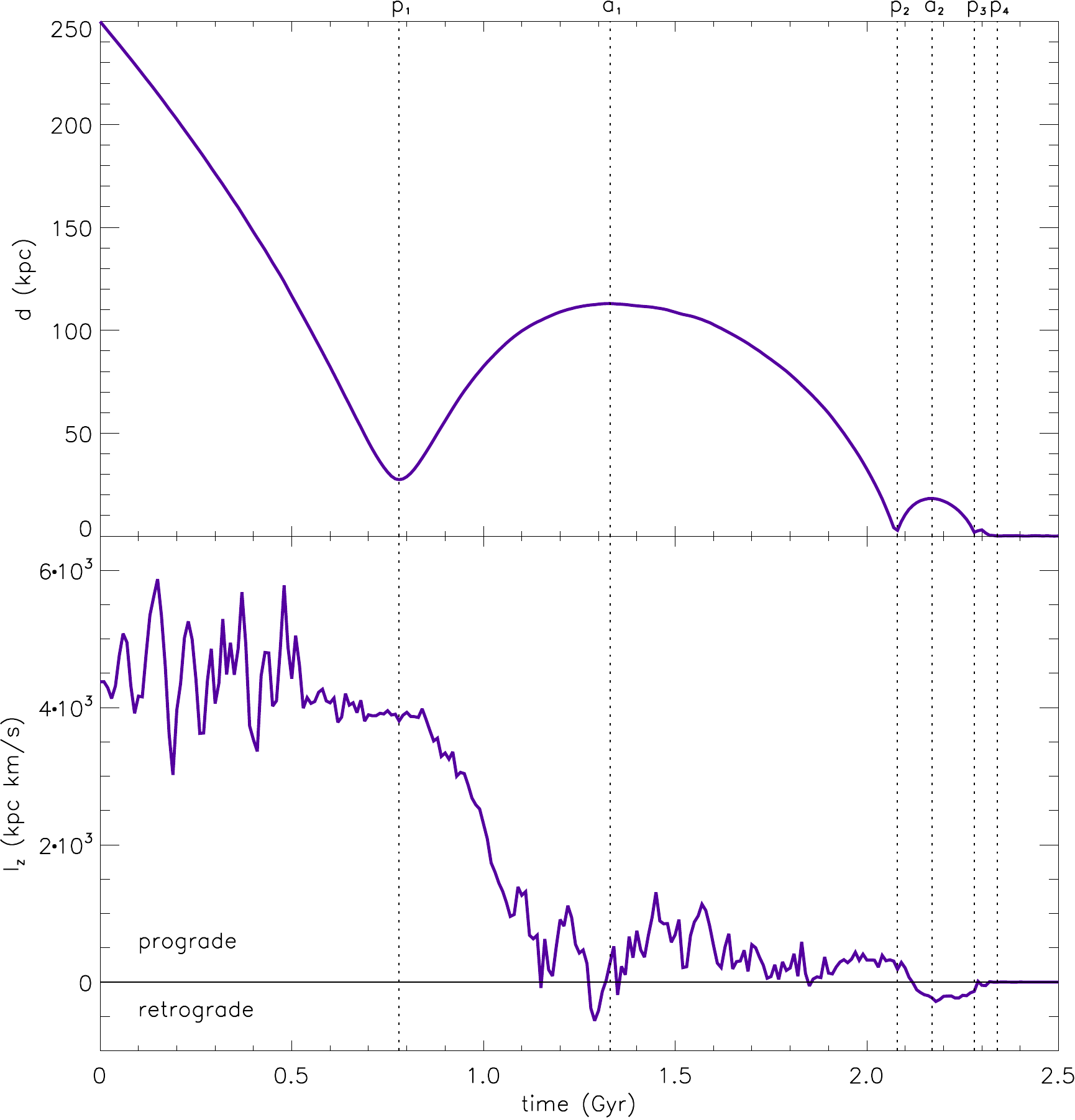}
 \caption{ Separation (d) and specific orbital angular momentum (\ensuremath{l_z}) of the two progenitor galaxies as a function of time. Times \ensuremath{p_{1}}, \ensuremath{p_{2}}, \ensuremath{p_{3}} and \ensuremath{p_{4}} indicate the time of the 1st, 2nd, 3d and 4th pericentric passage. Times \ensuremath{\alpha_{1}} and \ensuremath{\alpha_{2}} denote the 1st and the 2nd apocentric passage. The orbital spin of the two progenitors changes after \ensuremath{p_{2} }. }.
     \label{fig:figure3}
\end{figure}

One can clearly see from the velocity maps the presence of a large-scale KDC of radius $\sim$2 kpc in the center of the elliptical merger remnant (Figure~\ref{fig:figure2}).   This component is a counter-rotating component with respect to the outer body of the galaxy and is most prominent in the ``old" stellar population kinematics: stars that initially belonged to the two progenitor galaxies. 

On the other hand, ``young" stars form a stellar disk which is almost aligned with the orbital plane of the two progenitors. This young stellar disk is rotationally supported and strongly prograde-rotating, with a maximum velocity which is 4 times higher than the one of the ``old stars". We also note a weak sign of counter-rotation in the central region of the disk, seen in the extracted stellar rotation curve. 

Notably, stars that were initially part of the two progenitors' bulges (``Pr.Bulge") are globally counter-rotating in the merger remnant, exhibiting almost a solid-body rotation.

One can also see the presence of two symmetrical off-centered peaks in the ``all stars" stellar velocity dispersion map of Figure~\ref{fig:figure1}. This feature is commonly observed in ETGs with counter-rotating components (CRC). These galaxies are called ``$2\sigma$-galaxies"\citep{Krajnovic_2011}, and they were associated with external accretion of counter-rotating gas \citep{Rubin_1992} or major retrograde mergers \citep{Crocker_2009}. Here we see that a $2\sigma$-galaxy results from a single, prograde major merger. We also note that the $2\sigma$-feature is more prominent in the ``all stars" map, where the young disk of stars is present, and less strong in the ``old stars" map, even though the CRC is more prominent in the latter. In real ETGs a $2\sigma$-feature usually implicates the existence of a CRC, while the opposite is more ambiguous. Here we show that the $2\sigma$-feature might arise because of the presence of the CRC, but it is enhanced only if one of the components is fast-rotating\footnote{We should note, however, that most $2\sigma$-galaxies do not exhibit a centrally peaked velocity dispersion, like the one presented here.}. 

\begin{figure*}
\center
\includegraphics[trim = 2cm 9.1cm 3cm 4cm, clip, width=19cm]{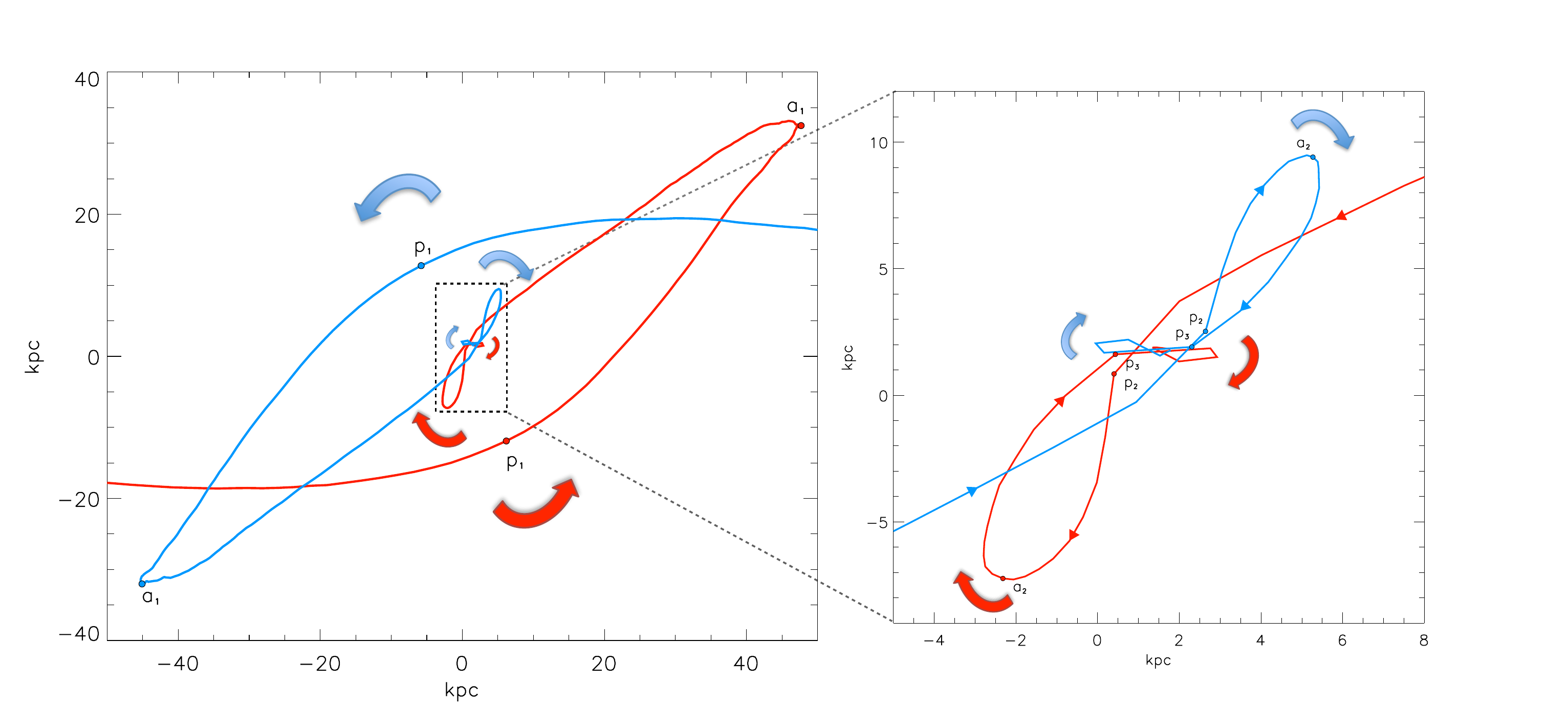}
\center
 \caption{Merging orbits of the two progenitor galaxies, with the orbital plane seen face-on.  Each curve (red and blue) corresponds to the orbit of each progenitor's bulge center of mass; the red curve denotes the galaxy that has its disk aligned with the orbital plane. Red and blue arrows show the direction of the orbital spin. \textbf{Left:} \ensuremath{p_{1}} denotes the 1st pericentric passage, when the two galaxies follow prograde orbits. The orbital spin changes sign right after the second pericentric passage \ensuremath{p_{2}}. \textbf{Right:} Zoom-in view of the region inside the dashed line on the right, when the orbits become retrograde.}
    \label{fig:figure4}
\end{figure*}
 
\section{ORIGIN of the Kinematic Decoupling}


In order to understand the origin of the kinematic decoupling in the central region of the galaxy, we study the behavior of the merging orbits of its two progenitors: Figure~\ref{fig:figure3} shows the separation (d) and the specific orbital angular momentum (\ensuremath{l_z}) for one of the two progenitors as a function of time. The merging orbits are shown in Figure~\ref{fig:figure4}, viewed face-on with respect to the initial orbital plane.

At the time the two progenitors reach their first pericenter (\ensuremath{p_{1}}=0.78 Gyr), they become tidally distorted, resulting in long trailing arms that expel loosely bound material from their disks. The orbital angular momentum is decreasing due to mass loss and dynamical friction (Figure~\ref{fig:figure3}). This causes the galaxies to approach their second pericenter (\ensuremath{p_{2}}=2.08 Gyr) with almost radial orbits (Figure~\ref{fig:figure4}). After their second encounter, the two galaxies change their orbital spin and follow a retrograde orbit for $\sim$300 Myr, until they finally reach their coalescence at t$\sim$2.4 Gyr.

\begin{figure*}
\center
\includegraphics[trim = 10mm 7cm 30cm 5mm, clip, width=15cm]{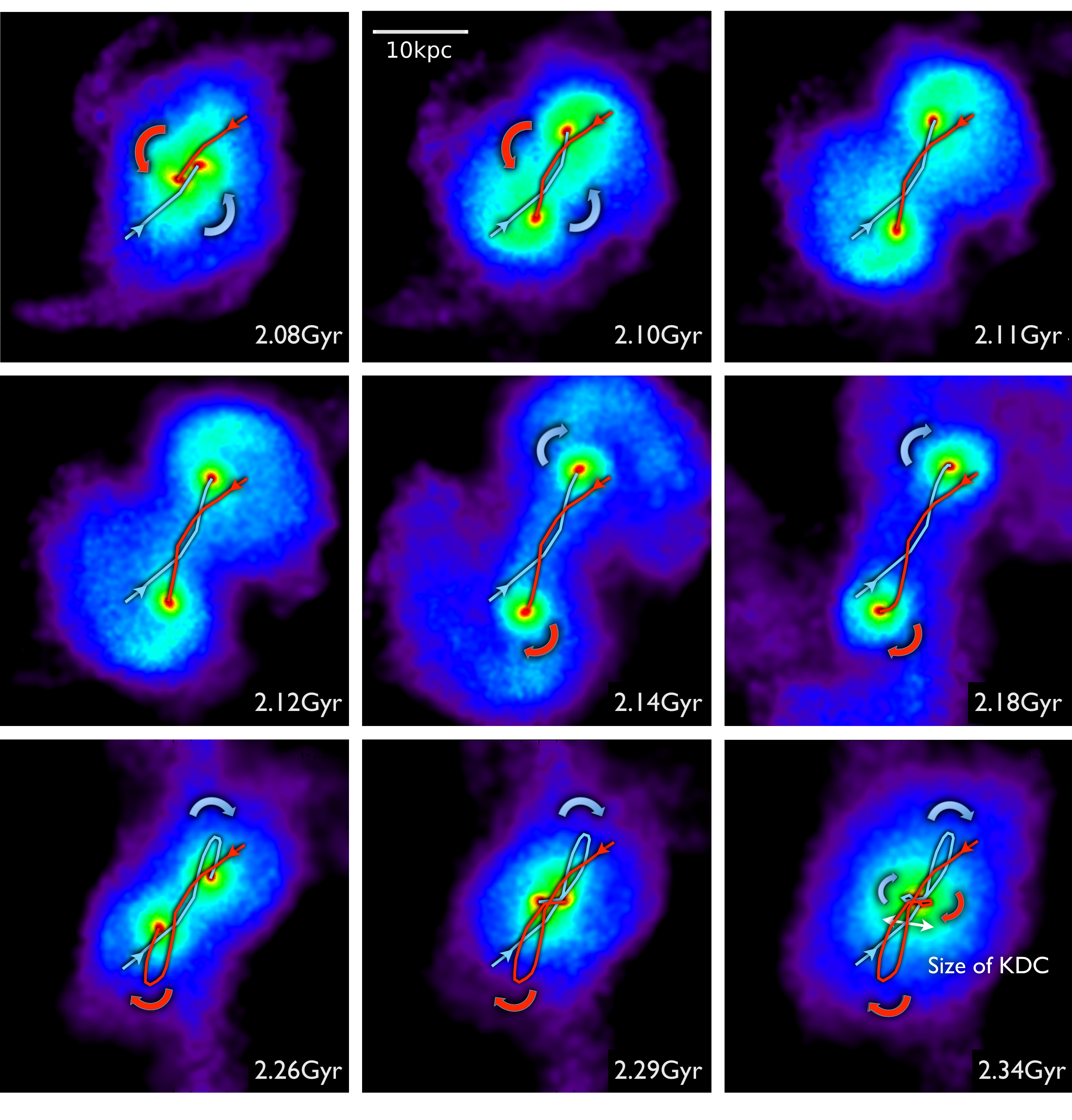}
\caption{Stellar surface density of the two progenitor galaxies after their second pericentric passage (2.08 Gyr) where the orbital spin changes sign. During this time, strong tidal interactions acting upon the two progenitors result in short-lived trailing spiral arms (2.08-2.18 Gyr), along which loosely bound material gets ejected from their main bodies. This results into a strong reactive force that causes the two galaxies to change their orbital spin and follow retrograde trajectories until their final coalescence at t$\sim2.4$ Gyr. The width of the last oscillation before coalescence corresponds to the size of the KDC in the final merger remnant. Images were created with SPLASH \citep{Price2007}. }
\label{fig:figure5}
\end{figure*}

The sudden change from a prograde to a retrograde merger can be understood in the framework of reactive forces.
Due to strong tidal interactions during the merger, the two progenitor galaxies are systems of variable mass; mass is constantly ejected along their short-lived trailing arms. The mass loss from each system results in a reactive force, known as the  \cite{Mestschersky_1902} force :
 \begin{equation}
 \overrightarrow{R}=\dot{m}(\overrightarrow{\upsilon}-\overrightarrow{V})
 \end{equation}
where $\dot{m}$ is the mass loss rate, $\overrightarrow{V}$ the bulk velocity of the system and $\overrightarrow{\upsilon}$ is the velocity of the outflowing matter. The Mestschersky force acts upon the two galaxies as a ``reactive thrust" which, if strong enough, can cause the change of the orbital spin.

Figure~\ref{fig:figure5} shows in detail this effect after the second pericentric passage. After the two progenitors approach closely, strong tidal forces that act upon them result in short-lived trailing arms, which mainly consist of their disks' stellar component. Loosely bound material gets ejected along these arms, resulting into a strong reactive thrust to the main bodies of the progenitors, causing them to change their orbital spin and follow retrograde trajectories until their coalescence at t$\sim$2.4 Gyr\footnote{The suggested mechanism could be responsible for the formation of KDCs in non-retrograde close encounters, e.g.\cite{Barnes_2002}.}.

The central region of the final remnant is counter-rotating and the width of the last oscillation before coalescence corresponds to the size of the KDC ($\sim$2 kpc), which is prominent in the center of the galaxy for more than 2 Gyr after the kinematic decoupling of its progenitors.

Under this framework, one can explain why stars that were initially part of the progenitors' bulges show global counter-rotation in the post-merger kinematics (Figure~\ref{fig:figure2}): these stars, more tightly bound in the centers of the galaxies during their close encounters, can track the behavior of the orbital spin of their progenitors' center of mass before coalescence.

On the other hand, the outer parts of the galaxy keep the initial prograde spin. Gas and stars ejected during the merger are subsequently re-accreted, while inheriting the outer prograde spin, forming the prograde-rotating outer part of the remnant. 

We suggest that the Mestschersky force is present in every stage of the merger. We interpret the change of sign of the orbital angular momentum near the first apocenter \ensuremath{\alpha_{1}}, as a result of this force (Figure~\ref{fig:figure3}), which is also seen as a change of curvature of the two merging orbits near \ensuremath{\alpha_{1}} in Figure~\ref{fig:figure4}. However, at this time in the merging process the effect is not strong enough to change the orbital spin.

\section{Summary and Discussion}
\label{Summary and Discussion}

We have shown that a KDC in an early-type galaxy can result from an initially prograde major merger of two disk galaxies. This finding is in contrast to the commonly suggested idea that KDC formation can only result from retrograde mergers. We show the plausibility of an orbital reversal of a prograde merger, caused by reactive forces that act upon the two progenitors due to mass loss, which results in KDC formation in the final merger remnant.

The KDC that resides in the center of the remnant shows strong counter-rotation for more than 2 Gyr after the final coalescence of its progenitors. The KDC is most prominent in the old stellar population of the galaxy (ages $\textgreater$ 5 Gyr) and is large in size (2 kpc radius), making it consistent with observations of KDCs in massive ETGs \citep{McDermid_2006} and comparable to the observed CRC/2-$\sigma$ galaxies \citep{Krajnovic_2011}. The fact that it results from an initially prograde merger provides a new channel for KDC formation that can add to the predicted rate of occurrence of KDCs and help towards explaining their observed fraction in ETGs.

The suggested formation scenario depends on reactive (Mestschersky) forces, that act upon the progenitors due to mass loss during the merger, causing the reversal of the orbital spin. Since prograde mergers result in substantial mass loss compared to retrograde mergers \citep{Toomre_1972, Barnes_1988}, we expect that such an effect is more likely to occur in prograde mergers. We would also expect this effect to depend on the mass ratio, the initial inclination, as well as the structural properties of the progenitor galaxies. Resolution effects might also influence properties of the KDC, such as its size and its position angle \citep[e.g.][]{Bois_2010}. 

Using the merger simulations presented in \cite{Moster_2011}, we note that the formation of the KDC does not depend on the particular form of feedback used or the specific values of hot and cold gas fractions employed in the progenitor galaxies.

A larger statistic of merger simulations will allow us to understand how common this new channel for  KDC formation is, which we plan to investigate in future work.


\acknowledgments
We acknowledge financial support to the DAGAL network from the People Programme (Marie Curie Actions) of the European Union's Seventh Framework Programme FP7/2007- 2013/ under REA grant agreement number PITN-GA-2011-289313. The numerical simulations used in this work were performed on the THEO cluster of the Max-Planck-Institut f{\"u}r Astronomie at the Rechenzentrum in Garching.


\bibliographystyle{apj}
\bibliography{ms}


\end{document}